\documentclass[twoside]{ilcws07}
\usepackage[latin1]{inputenc}
\usepackage[dvips]{graphicx,epsfig,color}
\usepackage{wrapfig,rotating}
\usepackage{amssymb,amsmath,array}

\begin{document}
\title{
A MAPS-based Digital Electromagnetic Calorimeter for the ILC} 
\author{J.~A.~Ballin, P.~D.~Dauncey, A.-M.~Magnan$^\dagger$, M.~Noy$^1$\\
	Y.~Mikami, O.~Miller, V.~Rajovi\'{c}, N.~K.~Watson, J.~A.~Wilson$^2$\\
	J.~P.~Crooks, M.~Stanitzki, K.~D.~Stefanov, R.~Turchetta,
      M.~Tyndel, E.~G.~Villani$^3$ 
\vspace{.3cm}\\
1 - Department of Physics, Imperial College London, London, UK\\
2 - School of Physics and Astronomy, University of Birmingham, Birmingham, UK\\
3 - Rutherford Appleton Laboratory, Chilton, Didcot, UK\\
$\dagger$ - Contact: {\tt A.Magnan@imperial.ac.uk}
}

\maketitle

\begin{abstract}
A novel design for a
silicon-tungsten electromagnetic calorimeter is described, based on Monolithic Active Pixel 
Sensors (MAPS). A test sensor with a pixel size of $50 \times 50\,\mu$m$^2$ has been 
fabricated in July 2007.
The simulation of the physical sensor is done using a detailed three-dimensional charge spread algorithm. 
Physics studies of the sensor are done including
a digitisation algorithm taking into account the charge sharing, charge collection efficiency, 
noise, and dead areas. The influence of the charge sharing effect is found to be
important
and hence 
needs to be measured precisely. 

\end{abstract}

\section{Introduction and pixel design}

The MAPS R\&D program is part of the CALICE~\cite{CALICE} collaboration and
proposes a swap-in solution to the existing analogue electromagnetic calorimeter
(ECAL) design~\cite{ECAL}, leaving the mechanical design unchanged. A first proof of concept sensor has been fabricated.\\
\begin{wrapfigure}{l}{0.4\columnwidth}
\centerline{\includegraphics[width=0.35\columnwidth]{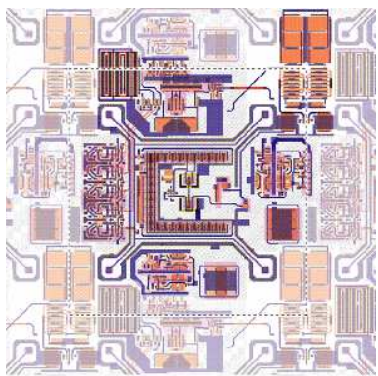}}
\caption{Example of sensor layout: the so-called presampler design.}\label{Fig:sensor}
\end{wrapfigure}
\indent The concept is to develop a digital electromagnetic calorimeter where
each pixel reports only a single bit. This requires a low probability
for multiple particles within a pixel, and that probability is reduced to an acceptable
level with a cell size of $50 \times 50\,\mu$m$^{2}$.\\
\indent The charge collection is done mainly by diffusion: four diodes placed near the 
corners of each pixel have been optimised in order to minimise the charge sharing 
between pixels while maximising the charge collection (see Section~\ref{sensorSimu}). 
To limit the charge sharing effect, the sensitive thickness of the silicon epitaxial layer has been set to $12\,\mu$m. The total silicon thickness remains $300\,\mu$m.\\
\indent The readout electronics are mainly inside the pixel. A column of 5 pixels every 42 
is however needed for the electronic logic, which accounts for around 11\% of dead area. 
Figure~\ref{Fig:sensor} shows an example of the pixel layout, where one pixel is 
represented inside the dash-lined area, and the four diodes can be seen on the four 
corners. The components inside the diodes form the analogue circuitry, whereas the 
comparator and logic are distributed around the edges. 
The diodes are N-well to P-substrate. The in-pixel N-wells needed for
the PMOS transistors would also collect charge, leading to losses in
signal. In order to minimise this effect, a novel 
process has been devised: the INMAPS process isolates the insensitive N-well region by 
screening it with a $1\,\mu$m thick deep P-well implant.\\
\indent The main challenge for a full calorimeter will be the power dissipation. 
The current test sensor has not been 
optimised at all in terms of power, and consumes on average $40\,\mu$W/mm$^2$, whereas 
the analogue CALICE design target is $1\,\mu$W/mm$^2$. This will be improved in the second design.

\section{Sensor simulation}
\label{sensorSimu}

\begin{wrapfigure}{l}{0.4\columnwidth}
\centerline{\includegraphics[width=0.35\columnwidth]{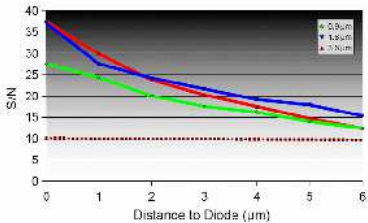}}
\caption{Sensor simulation: signal over noise ratio as a function of the distance of the input MIP to the diodes. The green, blue and red curves 
are respectively for 0.9, 1.8 and 3.6\,$\mu$m diodes.}
\label{Fig:diodes}
\end{wrapfigure}
The simulation of the charge collection of the sensor is done using Sentaurus TCAD~\cite{sentaurus}, a tool taking a 
precise description of the components in 3D.\\
\indent Figure~\ref{Fig:diodes} shows the signal over noise ratio as a function of the distance of the input MIP to the diodes, for three different diode sizes. The size of $1.8\,\mu$m seems appropriate to maximise the signal over noise ratio, while keeping the collected charge 
level to a reasonable level.\\
\indent Due to time constraints, results covering the whole pixel, with a $5\,\mu$m step in both 
direction, have been done using two approximations:
a pessimistic scenario, where no deep P-well is added, and the N-well is represented by a 
large central square collecting around 50\% of the total charge deposited in a pixel, and 
an optimistic scenario with a perfect deep p-well implant isolating completely the N-well.

\section{Physics simulation}
\label{physicsSimu}
The physics simulation of the whole detector using a Tesla-like design~\cite{LDC}, and a 
MAPS-based ECAL is done using GEANT4~\cite{geant4}. The distribution of the energy per 
hit for photon events is found to be stable from 500\,MeV up to 200\,GeV, 
confirming the assumption of having 1 MIP per cell on average. The energy resolution will then be given~by:\\
\centerline{$\sigma_{E}/E \propto \sqrt{\sigma_{N_{pixels}}^{2}+N_{noise}}/N_{pixels}$.}\\
\indent Due to their small size, the charge sharing between pixels is expected to be important, and will have a big influence on the number of hits above threshold, and hence on the energy 
measurements. Understand this phenomenon will be crucial, as well as a precise measurement 
of its effects, to prove the validity of such a calorimeter.\\

\begin{figure}[h!]
\begin{minipage}[h]{0.45\columnwidth}
\centerline{\includegraphics[width=0.9\columnwidth]{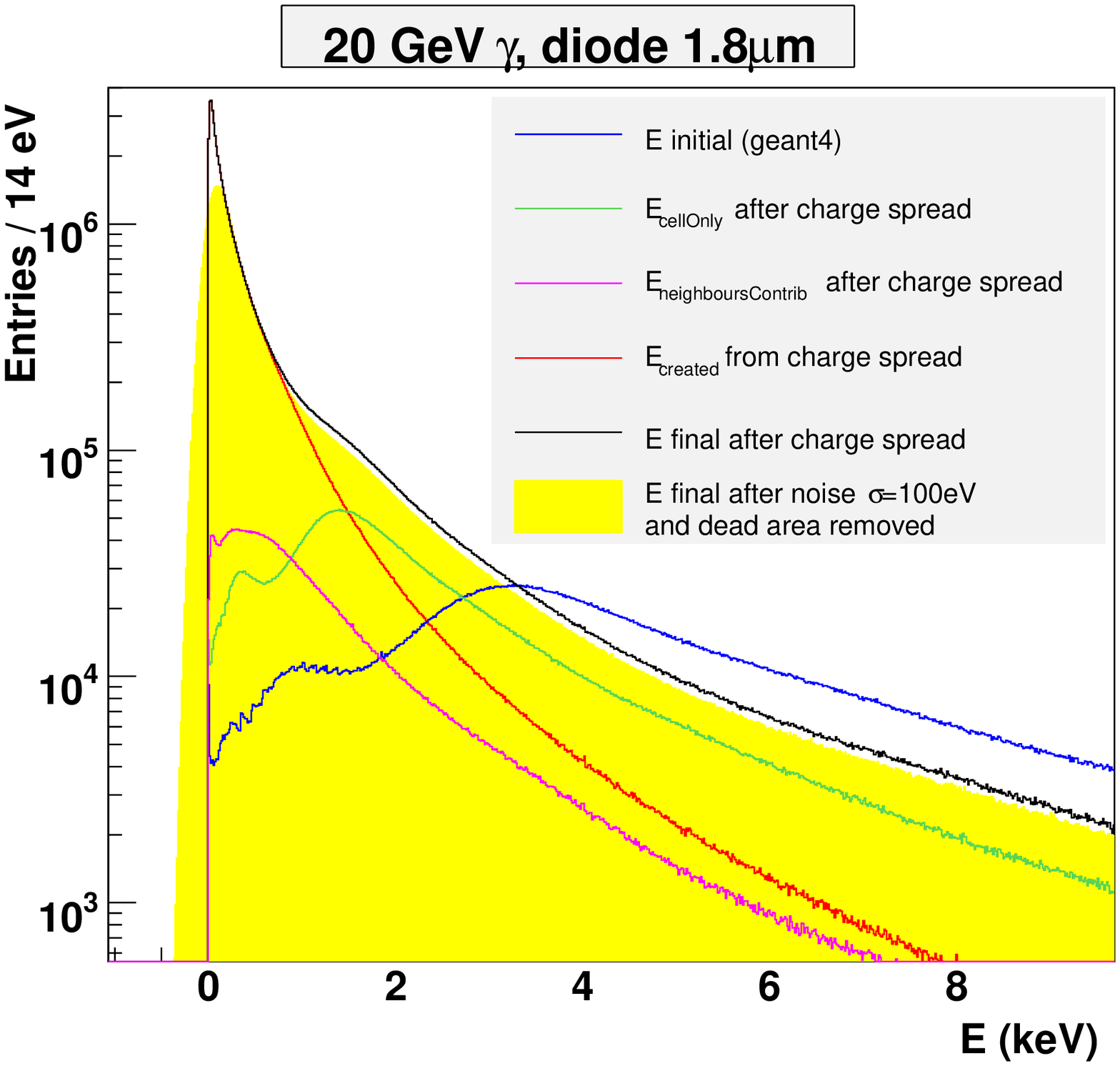}}
\caption{Energy per hit for the different digitisation steps.}\label{Fig:digiSteps}
\end{minipage}
\hfill
\begin{minipage}[h]{0.45\columnwidth}
\centerline{\includegraphics[width=0.9\columnwidth]{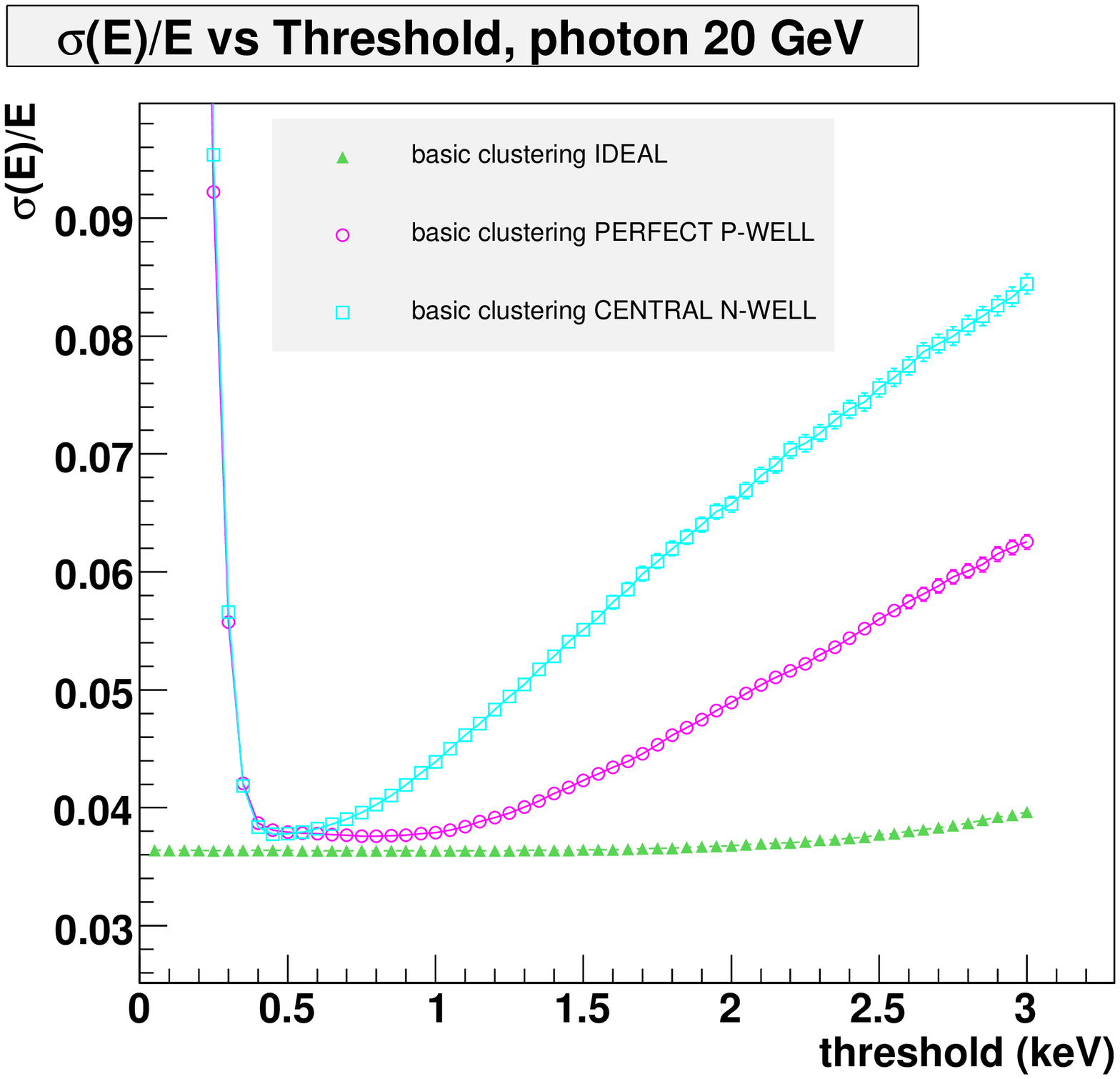}}
\caption{Energy resolution versus threshold before digitisation (``IDEAL'') and for the two charge spread models.}\label{Fig:EresoModel}
\end{minipage}
\end{figure}

\indent The digitisation, required for a realistic simulation, is done in several steps, with results displayed in Figure~\ref{Fig:digiSteps} 
in terms of energy per hit, where the charge spread model assumes a perfect
p-well.
The simulation of the energy deposited in $5 \times 5\,\mu$m$^{2}$ cells is done with the Mokka~\cite{mokka} application. The charge sharing results from the sensor 
simulation (see Section~\ref{sensorSimu}) are then applied, giving for each deposition the 
percentage of energy seen by the pixel and by its eight closest neighbours.
The results before applying the charge spread are displayed in dark blue.
The other coloured curves show various intermediate contributions,
and the black curve is the result per 
pixel when the different contributions have been summed.
A noise of $\sigma = 100$\,eV (which corresponds to 30 electrons, to be cross-checked with the sensor test setup) is then
added.
The influence of 
the noise on the energy resolution is found negligible for a threshold value above 600\,eV, or 
$6\sigma$ of the noise. This is close to the region where the energy resolution is found minimal 
(see Figure~\ref{Fig:EresoModel}).
Dead areas of 5 pixels every 42 pixels are removed, giving the yellow histogram, Figure~\ref{Fig:digiSteps}. 
This was found to degrade the energy resolution measurement by 6\%, for single photon events at 20\,GeV.
A basic MIP clustering algorithm is finally applied, according to the
number of closest neighbours, and gives a 16\% improvement when 
calculating the energy resolution versus threshold, for a 20 GeV photon.

\indent Figure~\ref{Fig:EresoModel} displays the energy resolution versus threshold for 20\,GeV
photons for the two extreme 
models we have taken into account up to now. From these curves, two very positive remarks 
can be drawn. Firstly, the value of the threshold corresponding to the minimum in terms of energy resolution 
lies outside the noise region, i.e. above 5$\sigma$. Secondly, the minimum value stays constant over
a range of threshold settings
and is close to the value found when no charge spread is assumed. In the pessimistic N-well
scenario, the minimum region is quite narrow, whereas the optimistic scenario predicts a flat region 
between 5 and 10$\sigma$ of the noise. The reality is expected to lie in between. The influence of 
the charge spread model is therefore crucial, and hence needs to be measured precisely, and compared 
with the sensor simulation results.

\section{Conclusion}

The sensor test setup is now completed, with several designs received from the foundry beginning of 
July 2007. 
Simulation shows a MAPS-based calorimeter has the potential to give good energy resolution but
several aspects of the sensor response must be measured to cross-check the simulation.







\begin{footnotesize}



%

\end{footnotesize}


\end{document}